\documentclass[conference,10pt]{IEEEtran}
\IEEEoverridecommandlockouts
\usepackage[T1]{fontenc}
\usepackage{type1cm}
\usepackage{url}
\usepackage{fix-cm}
\usepackage{graphicx,color,comment}
\usepackage[table]{xcolor}
\usepackage{amsmath, amsthm, amsfonts, amssymb, amsbsy,nccmath}

\usepackage{enumerate}
\usepackage{lipsum}
\newtheorem{lemma}{Lemma}
\usepackage{textgreek}
\usepackage{comment}
\usepackage{textcomp}
\usepackage{tikz}
\usetikzlibrary{positioning, fit, backgrounds, arrows.meta, calc, shapes.geometric, decorations.pathreplacing}
\usepackage{booktabs}
\usetikzlibrary{positioning, arrows.meta, shapes.geometric, calc, decorations.pathreplacing}
\usepackage[ruled,vlined,linesnumbered]{algorithm2e}
 \newtheorem{proposition}{Proposition}
\usepackage[most]{tcolorbox}
\usepackage[T1]{fontenc}
\usepackage{trace}% optional, ignore

\usepackage[sort,compress]{cite}
\usepackage{epsfig}
\usepackage{epstopdf}
\usepackage{mathtools}
\usepackage{dsfont}
\usepackage{cleveref}
\usepackage{subfig}
\usepackage{stfloats}
\usepackage{fontawesome5}

\newlength{\sciH}
\newcommand{\scicolorbar}[1]{% #1 = bar height
  \setlength{\sciH}{#1}%
  \begin{tikzpicture}[font=\scriptsize, x=1cm, y=\sciH]
    \definecolor{jdb}{rgb}{0,0,0.5}\definecolor{jb}{rgb}{0,0,1}
    \definecolor{jc}{rgb}{0,1,1}\definecolor{jy}{rgb}{1,1,0}
    \definecolor{jr}{rgb}{1,0,0}\definecolor{jdr}{rgb}{0.5,0,0}
    \shade[bottom color=jdb, top color=jb] (0,0)     rectangle (0.35,0.125);
    \shade[bottom color=jb,  top color=jc] (0,0.125) rectangle (0.35,0.375);
    \shade[bottom color=jc,  top color=jy] (0,0.375) rectangle (0.35,0.625);
    \shade[bottom color=jy,  top color=jr] (0,0.625) rectangle (0.35,0.875);
    \shade[bottom color=jr,  top color=jdr](0,0.875) rectangle (0.35,1);
    \draw (0,0) rectangle (0.35,1);
    \node[right] at (0.35,0) {$0$};
    \node[right] at (0.35,1) {$1$};
    \node[rotate=90] at (-0.2,0.5) {Criticality};
  \end{tikzpicture}}

\theoremstyle{definition}
\newtheorem{definition}{Definition}[section]

\usepackage[inline]{enumitem}   
\makeatletter
\newcommand{\inlineitem}[1][]{%
\ifnum\enit@type=\tw@
    {\descriptionlabel{#1}}
  \hspace{\labelsep}%
\else
  \ifnum\enit@type=\z@
       \refstepcounter{\@listctr}\fi
    \quad\@itemlabel\hspace{\labelsep}%
\fi}
\makeatother

\usepackage{float}
\floatstyle{plaintop}
\restylefloat{table}

\makeatletter
\newcommand\fs@spaceruled{\def\@fs@cfont{\bfseries}\let\@fs@capt\floatc@ruled
  \def\@fs@pre{\vspace{0.5\baselineskip}\hrule height.8pt depth0pt \kern2pt}%
  \def\@fs@post{\kern1pt\hrule\relax}%
  \def\@fs@mid{\kern2pt\hrule\kern2pt}%
  \let\@fs@iftopcapt\iftrue}
  \newcommand\fs@betterruled{%
  \def\@fs@cfont{\bfseries}\let\@fs@capt\floatc@ruled
  \def\@fs@pre{\vspace*{5pt}\hrule height.8pt depth0pt \kern2pt}%
  \def\@fs@post{\kern2pt\hrule\relax}%
  \def\@fs@mid{\kern2pt\hrule\kern2pt}%
  \let\@fs@iftopcapt\iftrue}
\makeatother

\usepackage{footnote}

\graphicspath{{Figures/}}

\begin{document}

\title{STAR: Semantic-Criticality Tiered Allocation of Radio Resources via Zero-Shot Learning\vspace{-3mm}}

\vspace{-2mm}\author{
\IEEEauthorblockN{
Albert Shaju\IEEEauthorrefmark{1},
Christo Kurisummoottil Thomas\IEEEauthorrefmark{1}, and Mingzhe Chen\IEEEauthorrefmark{2}
}
\IEEEauthorblockA{
\IEEEauthorrefmark{1}Department of Electrical and Computer Engineering,
Worcester Polytechnic Institute, Worcester, MA, USA,\\
\IEEEauthorrefmark{2}Department of Electrical and Computer Engineering, University of Miami, Coral Gables, FL, USA,\\
Emails: \{ashaju@wpi.edu, cthomas2@wpi.edu, mingzhe.chen@miami.edu\}
\vspace{-3mm}}
\thanks{\vspace{-4mm}This research was supported by the AI-RAN Alliance Innovation Fund.}\vspace{-6mm}}
\maketitle

\vspace{-5mm}\begin{abstract}
In remote inference, edge devices transmit high-resolution images over wireless links to a receiver, where an artificial intelligence (AI) model answers task-specific questions about these images to enable reliable decisions. These answers are often determined by a small set of \emph{task-causal} pixels, such as those of text, whose corruption changes the answer, so these pixels must be delivered with high fidelity. Conventional transmission protects every pixel equally, so it either consumes many resource elements (REs) or corrupts these regions at low signal-to-noise ratio (SNR). To overcome these limitations, in this paper, a zero-shot, domain-agnostic framework, termed semantic-criticality tiered allocation of resources (STAR), is proposed. In particular, a per-pixel semantic-criticality indicator (SCI) is derived from the attention of a frozen vision-language model, so new tasks and domains are handled by changing only the prompt, without retraining. The resulting task causality-aware resource allocation problem over protection tiers is solved exactly by a threshold rule on the SCI, realized by a coded tier for the task-causal pixels and an uncoded, downsampled base layer for the full image. Simulations over a 5G New Radio link demonstrate that STAR attains a semantic accuracy of at least $0.98$ for $E_s/N_0\ge-4$~dB and, at comparable semantic reliability, reduces the REs of robust coded transmission by $59.5\%$ and $91.5\%$ on factory- and vehicle-monitoring datasets, respectively, the latter without any change to the models. Although a deep joint source-channel coding baseline uses fewer REs at comparable semantic accuracy, it must be retrained with labeled data for every new task or domain.
\end{abstract}

\vspace{-2mm}\section{Introduction}\vspace{-1mm}
Deep learning based semantic communication (SC), called deep joint source-channel coding (DeepJSCC) \cite{bourtsoulatze2019deep} and \cite{xie2021deep}, learns the encoder and decoder jointly for a given task, source distribution, and channel. These systems depend on \emph{task-specific datasets and offline training}, so they generalize poorly and must be retrained for each new task or domain. Moreover, replacing the source-channel coding chain with a learned transceiver also requires changes across multiple layers of the communication stack, which raises \emph{interoperability challenges} with existing cellular standards.

\vspace{-2mm}\subsection{Related Work}\vspace{-2mm}

There has been a surge of recent works on generative artificial intelligence (AI) based SC schemes that transmit a compact description of the image, such as a semantic map \cite{grassucci2023generative,Strinati2024GOALS}, a compressed latent \cite{guo2024diffusion}, or text and multi-modal prompts \cite{nam2024language,du2023generative}, from which a diffusion or text-to-image model regenerates the image at the receiver (RX). When this description is extracted by an encoder trained on a specific dataset, objects in out-of-distribution (OOD) inputs can be misidentified, leading to a semantically distorted image at the RX. %This is avoided either by building the system from off-the-shelf foundation models without any training \cite{nam2024language}, or by falling back to a multimodal large language model (MLLM) when the confidence of the trained encoder is low and refining the regenerated image through a generate--criticize loop between two MLLMs \cite{zhang2024addressing}. %In both cases, however, what is transmitted does not depend on the task at the receiver. %While these task-oriented approaches reduce the communication overhead, they discard the raw data at the transmitter, so a \emph{hallucination} introduced during extraction or generation \cite{li-etal-2023-evaluating} yields a plausible output that the receiver has no reference to detect. This failure is critical in safety monitoring, where a missed hazard, such as an active fire, leaves the receiver with no information from which to detect the error or initiate a response. 
%Limiting such errors for out-of-distribution inputs has relied on additional multimodal large language models (MLLMs) at both ends, with a generate-criticize loop at the receiver \cite{zhang2024addressing}.
On the other hand, importance-aware schemes \cite{guo2023semantic,park2024vision,christopoulos2000efficient,liu2024ofdm} keep the original data verifiable by transmitting it in full, and they reduce resource use by allocating resources according to the \emph{semantic importance} of each part. Pre-trained language models have been used to quantify the semantic importance of text frames and allocate transmit power accordingly \cite{guo2023semantic}, and the class-token attention of a vision transformer has been used to assign quantization bits to image patches over a binary symmetric channel \cite{park2024vision}. However, the attention map in \cite{park2024vision} requires fine-tuning the transformer on a labeled classification dataset, so whenever the task or the environment changes during operation, new labeled data must be collected and the model fine-tuned again, which  \emph{hinders online adaptation}. Region-of-interest coding \cite{christopoulos2000efficient} provides similar prioritization with regions given by a fixed rule or detector. In these schemes, importance is either fixed regardless of the RX task \cite{guo2023semantic,christopoulos2000efficient}, so it cannot prioritize the content that a given task depends on, or tied to a single task through fine-tuning \cite{park2024vision}, so \emph{a new task requires new labeled data and retraining}.  Moreover, the tolerance of semantic content to channel errors varies with its spatial redundancy, since semantic concepts spread over many pixels remains recognizable at reduced fidelity whereas content concentrated in a few pixels, such as text, is corrupted by a few erroneous pixels, and we refer to this variation as \emph{tiered semantic criticality}. 
To overcome these limitations, we build on foundation models, whose large-scale pre-training on diverse data allows them to perform many different tasks from a natural-language instruction alone, without updating their weights, a capability known as \emph{zero-shot in-context learning}\cite{brown2020gpt3}.

\vspace{-3mm}\subsection{Our Contributions}\vspace{-2mm}
The main contribution of this paper is \emph{Semantic-criticality Tiered Allocation of radio Resources (STAR)} via zero-shot learning with foundation models. In particular, STAR can generalize to new tasks and domains by a simple change in prompts, and it concentrates channel protection on the task-causal pixels of each image while still delivering the full image for audits and incident analysis. Specifically, we introduce the \emph{semantic-criticality indicator (SCI)}, a per-pixel map that estimates the causal influence of each pixel on the task from the attention of a frozen vision-language model (VLM). Furthermore, we formulate the \emph{causality-aware resource allocation} problem over per-pixel protection tiers and show that its exact solution is a threshold rule on the SCI map, which we realize with a two-layer transmission scheme on the existing modulation and coding chain, where a low-cost base layer carries the whole image and a robust layer carries only the sparse task-causal pixels. Over an orthogonal frequency division multiplexing (OFDM) link with the 3GPP tapped delay line A (TDL-A) channel \cite{3gpp38901}, STAR matches the semantic reliability of binary phase shift keying (BPSK) with a rate-$1/3$ convolutional code on the whole image while using $59.5\%$ fewer resource elements (REs) on a synthetic factory-monitoring dataset and $91.5\%$ fewer on real vehicle-monitoring photographs, the latter with no change to the models or parameters, which demonstrates domain-agnostic operation.

\vspace{-2mm}\section{System Model and Problem Formulation}
\vspace{-1mm}

Consider a set of edge devices, such as patrol robots and fixed cameras, that periodically capture images of $H$ rows and $W$ columns of pixels and transmit them over the cellular uplink to a remote control center that performs \emph{remote inference} on the received images to trigger actions. Each action depends on a task $q\in\mathcal{Q}$, expressed as a natural-language question about the image, and we assume that the action is triggered correctly if remote inference answers its question correctly on the received image. \emph{The control center also stores the received images for audits and incident analysis, so it requires the images themselves and not only the extracted answers}.

As an illustrative example, we consider a smart factory whose equipment rooms are monitored by patrol robots and fixed cameras. Fig.~\ref{fig:scene-clean} shows an equipment room in normal operation. Each machine carries a status panel, enlarged in Fig.~\ref{fig:scene-zoom}, that displays its equipment status, pressure, temperature, and power source, and a room may contain a maintenance technician, as in Fig.~\ref{fig:scene-tech}, or an active fire, as in Fig.~\ref{fig:scene-fire}. Typical questions are ``What is the temperature reading?'' and ``Is there a fire?'', which trigger actions such as stopping an overheating machine, raising a fire alarm, or issuing a service call.

For a given set of tasks, only some pixels of an image determine the answers, in the sense that replacing them with a degraded version changes an answer, whereas the remaining pixels can be degraded without changing any answer. We refer to these as the \emph{task-causal} and \emph{task-invariant} pixels, respectively, which are defined formally in Definition~\ref{def:causal}. 
Since the devices share 5G NR links to the control center, each image should consume as few OFDM REs as possible while every action is still triggered correctly. Conventional transmission protects every pixel equally, so it either renders the task-causal pixels unreadable at low signal-to-noise ratio (SNR) or spends many REs. \emph{Our objective is to exploit the sparsity of the task-causal pixels to attain the same semantic reliability with fewer REs.}

\begin{comment}
    For a given set of tasks, only some regions of an image carry the information that the tasks depend on. We refer to the pixels in these regions as \emph{critical pixels}, and to the remaining pixels as normal pixels. The critical pixels are sparse. In the datasets of Section~\ref{sec:datasets}, they form $3.6\%$ of the pixels in the vehicle-monitoring dataset and $18.9\%$ in the factory-monitoring dataset, although the latter evaluates $7$ tasks per image.

The transmitters of all factories send their data to the control center over 5G NR links, which use OFDM. The channel resources of an OFDM link are divided into REs, which are scarce because the traffic of many factories converges at a single control center. The goal is therefore to minimize the REs per image while every action is still triggered correctly, which we refer to as high semantic reliability. Conventional transmission treats the data as bits without semantics, so it allocates the same channel resources to every pixel. Lossy compression with high-order modulation uses few REs, but it renders the critical pixels unreadable at low signal-to-noise ratio (SNR). Uncompressed transmission with low-order modulation and channel coding attains high semantic reliability, but it consumes many REs. STAR exploits the sparsity of the critical pixels to attain the same semantic reliability with fewer REs.

\end{comment}

\vspace{-2mm}\subsection{Problem Formulation}\label{sec:pf}\vspace{-2mm}

We consider that each of the $P\!=\!HW$ pixels has three color channels of $b$ bits, so it carries $B\!=\!3b$ bits. The task-invariant pixels are delivered by a base layer, which carries the whole image at reduced fidelity, decimated by a factor $D$ and mapped uncoded onto a constellation with $m_{b}$ bits per symbol. The task-causal pixels are delivered by protection tiers, where tier $k$ carries their full-resolution values with a rate-$r_k$ code and $m_k$ bits per symbol, and these values replace the base-layer values at the RX. Each protected pixel also carries its raster position $\mathrm{pos}_p$, encoded with $B_a=\lceil\log_2 P\rceil$ bits, so that the RX can place it. Let $x_p\in\{0,1,\dots,K-1\}$ be the tier of pixel $p$, where $x_p=0$ means that the pixel is carried by the base layer only. The number of REs per image is

\vspace{-3mm}\begin{equation}
J(\boldsymbol{x})=\frac{PB}{D^2 m_b}+\sum_{p=1}^{P}c_{x_p},\qquad
c_0=0,\;\; c_k=\frac{B+B_a}{m_k\,r_k}.
\label{eq:cost}
\vspace{-2mm}\end{equation}
Each tier $k$ therefore corresponds to a modulation and coding scheme (MCS) $(m_k,r_k)$ from an existing MCS set, carried as a separate transport block, so STAR requires no new physical-layer operation. %Each tier $k$ therefore corresponds to a modulation and coding scheme (MCS) $(m_k,r_k)$ with spectral efficiency $m_kr_k$ information bits per RE, and $c_k$ is the number of REs spent on a protected pixel.  Assigning a pixel to a tier is thus a per-pixel MCS selection from an existing MCS set, and each tier can be carried as a separate transport block with its own MCS, so STAR requires no new physical-layer operation.

Let $\epsilon_k(\gamma)$ denote the expected squared error per color channel, relative to the original $b$-bit value, of a pixel received through tier $k$ at $\gamma=E_s/N_0$, with $\epsilon_0$ for the base layer. The error of tiers $k\ge1$ is due only to the bit errors that remain after channel decoding, whereas the error of the base layer also includes the decimation error. Without loss of generality, we order the tiers so that $c_1<\dots<c_{K-1}$ and $\epsilon_0>\epsilon_1>\dots>\epsilon_{K-1}$.
\vspace{-2mm}\begin{lemma}\label{lem:dist}
If the bits of a $b$-bit color value are equiprobable and are flipped independently with probability $P_b$, the expected squared error of the value is $\frac{4^b-1}{3}P_b$. Hence
\vspace{-1mm}\begin{equation}
\epsilon_0(\gamma)\approx\epsilon_{\mathrm{dec}}(D)+\tfrac{4^b-1}{3}P_b^{(0)}(\gamma),\quad
\epsilon_k(\gamma)\approx\tfrac{4^b-1}{3}P_b^{(k)}(\gamma),
\label{eq:eps}
\end{equation}
where $\epsilon_{\mathrm{dec}}(D)$ is the interpolation error of decimation, $P_b^{(0)}$ is the uncoded bit error rate of the base constellation, and $P_b^{(k)}$ is the decoded bit error rate of tier $k$.
\vspace{-2mm}\end{lemma}
\begin{proof}
A flip of bit $i$ contributes $(2^i)^2=4^i$ to the squared error. \begin{figure}[t]
  \centering
  \subfloat[Normal operating event]{%
    \includegraphics[width=0.45\linewidth]{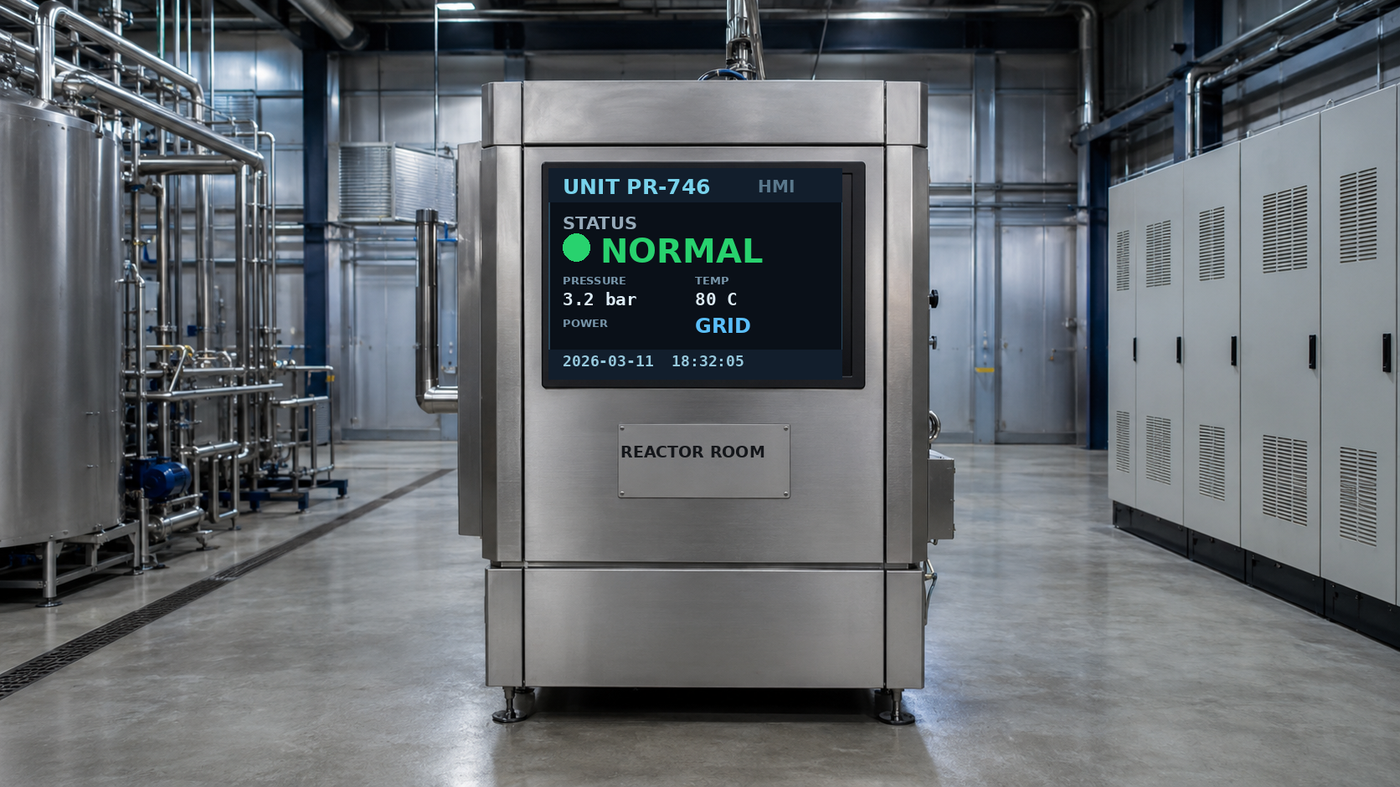}%
    \label{fig:scene-clean}}
  \hfil
  \subfloat[Status panel (enlarged)]{%
    \vspace{-3mm}\includegraphics[width=0.36\linewidth]{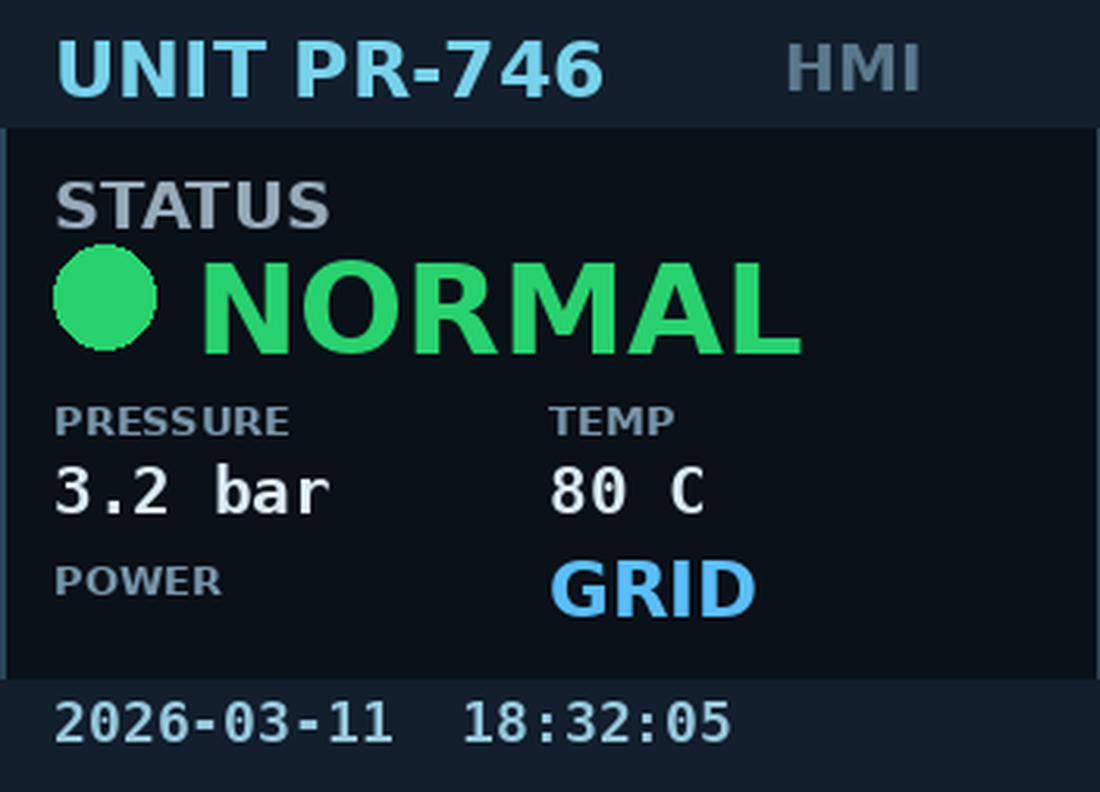}%
    \label{fig:scene-zoom}}
  \\[-3mm]
  \subfloat[Technician event]{%
    \vspace{-3mm}\includegraphics[width=0.45\linewidth]{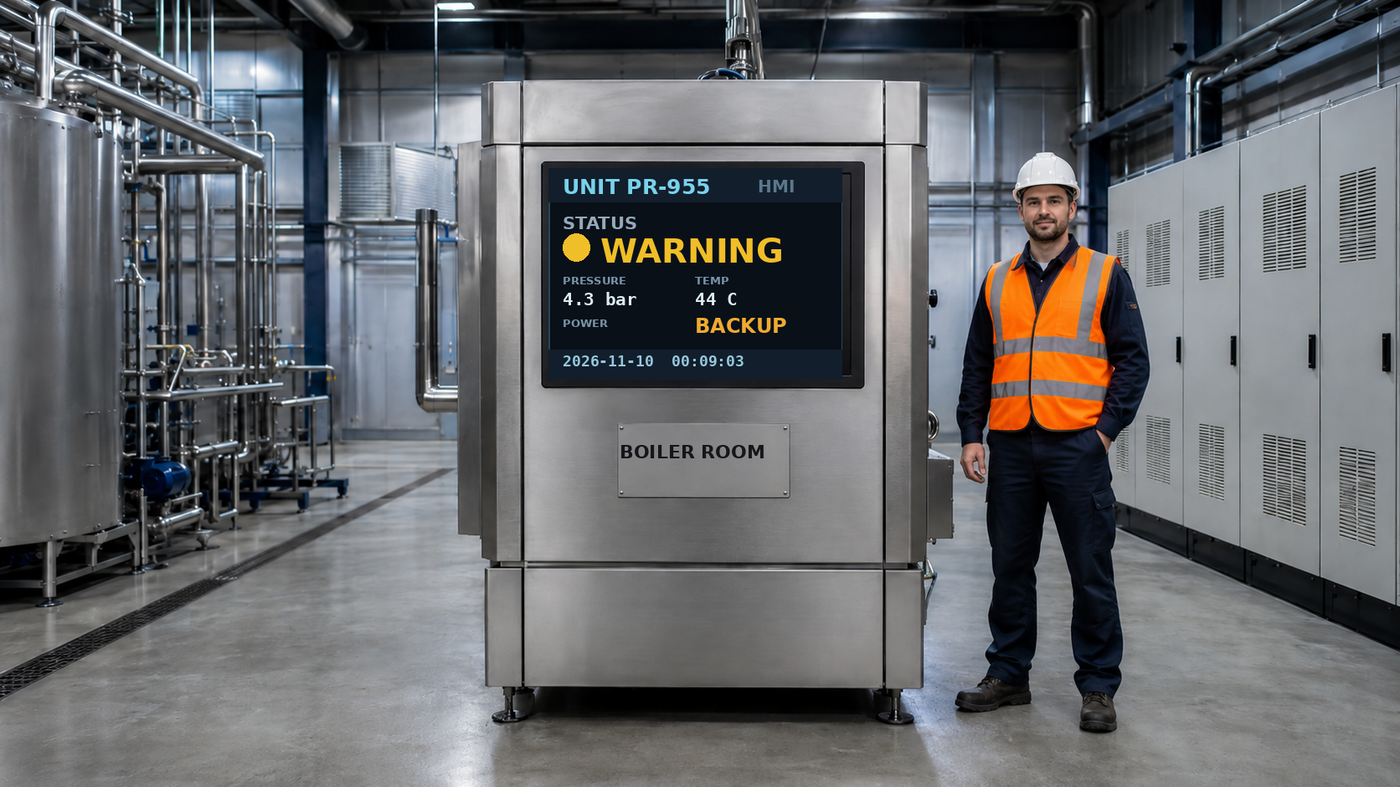}%
    \label{fig:scene-tech}}
  \hfil
  \subfloat[Fire event]{%
    \vspace{-3mm}\includegraphics[width=0.45\linewidth]{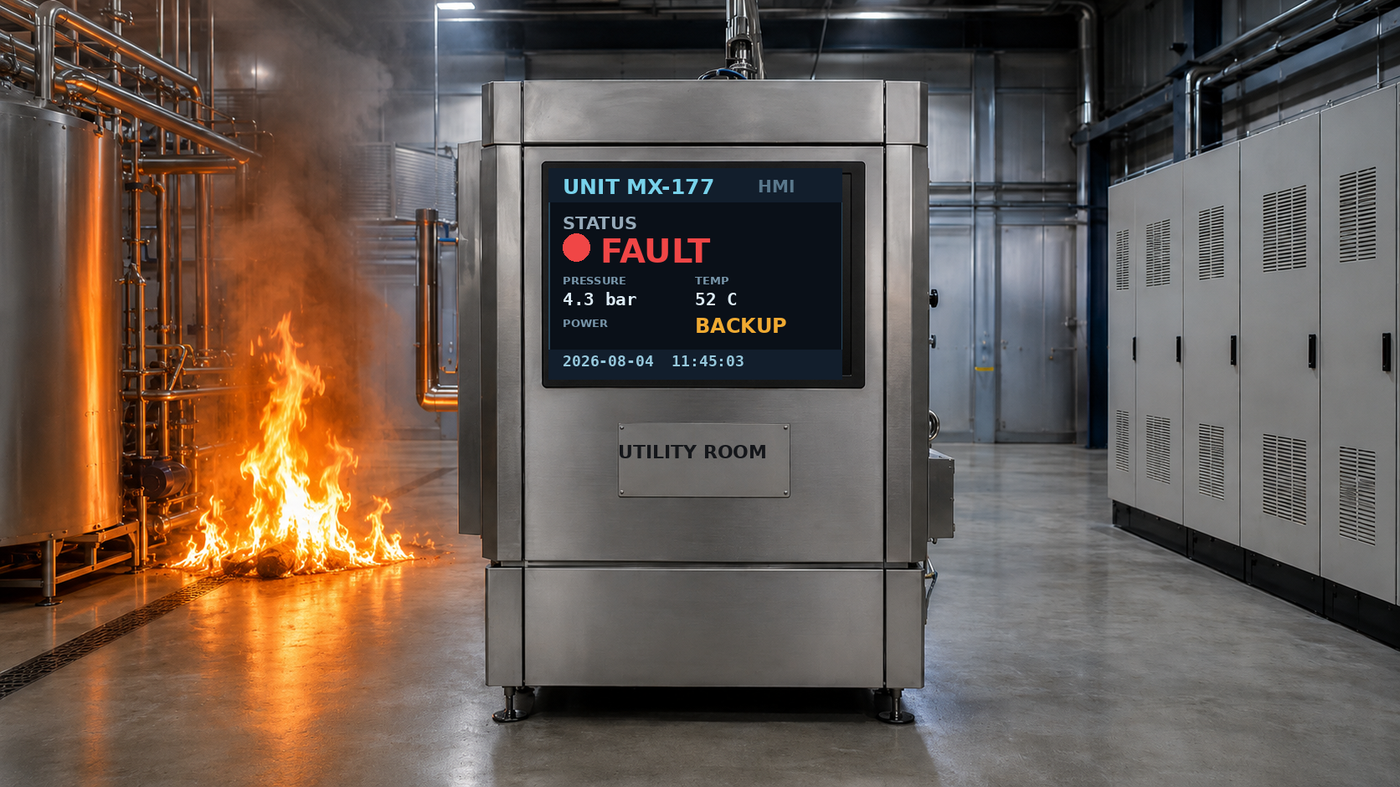}%
    \label{fig:scene-fire}}
  \vspace{-2mm}\caption{\small Sample images from a factory-monitoring environment.}
  \label{fig:dataset_scenes}
\vspace{-0mm}\end{figure}The signs of flips in different bits are independent and equiprobable, so the cross terms vanish in expectation, and the expected squared error is $\sum_{i=0}^{b-1}P_b4^i=\frac{4^b-1}{3}P_b$.
\vspace{-2mm}\end{proof}

For a task $q\in\mathcal{Q}$, let $f_q(I)$ denote the answer to $q$ on image $I$, and let $I_{\mathcal{R}\leftarrow\tilde I}$ denote the image in which the pixels of a set $\mathcal{R}$ are replaced by their base-layer reconstruction $\tilde I$.
\vspace{-2mm}\begin{definition}\label{def:causal}
The \emph{task-invariant component} of $I$ with respect to $q$ is a largest pixel set $\mathcal{R}_q$ such that $\Pr\big[f_q(I_{\mathcal{R}_q\leftarrow\tilde I})\neq f_q(I)\big]\le\varepsilon_q$, where the probability is over channel realizations. Its complement $\mathcal{C}_q$ is the task-causal component.
\vspace{-2mm}\end{definition}
The base layer thus acts as an intervention on every pixel that it carries alone. The answer to $q$ is invariant to this intervention on $\mathcal{R}_q$, whereas an intervention on $\mathcal{C}_q$ changes it. To quantify this per pixel, let $a_q(p)\in[0,1]$ be the \emph{causal influence} of pixel $p$ on the answer to $q$, that is, \emph{the normalized interventional effect} of pixel $p$ on $f_q$. It is small on $\mathcal{R}_q$ and large on $\mathcal{C}_q$, and its value within $\mathcal{C}_q$ grades how strongly each pixel must be protected. Further, we define the \emph{semantic distortion} of task $q$ as the largest causal influence-weighted error over the image for a specific channel quality $\gamma$,
\vspace{-1mm}\begin{equation}
\mathsf{D}_q(\boldsymbol{x},\gamma)=\max_{p}\,a_q(p)\,\epsilon_{x_p}(\gamma).
\label{eq:sd}
\vspace{-2mm}\end{equation}
This worst-case form reflects that corrupting a few highly influential pixels, such as those of a short text, can be enough to change the answer. We assume that the failure probability of task $q$ is non-decreasing in $\mathsf{D}_q$, so a reliability target for task $q$ is met by $\mathsf{D}_q\le\delta_q$. The tolerance $\delta_q$ encodes the criticality of task $q$, with a smaller $\delta_q$ for a more critical task, so the tasks may take any number of criticality levels. %Because the semantic distortion \eqref{eq:sd} weights the channel-induced error of each pixel by its interventional effect on the answer, the constraint $\mathsf{D}_q\le\delta_q$ tolerates channel errors on the task-invariant pixels and bounds them on the task-causal pixels. 
We therefore formulate the \emph{task causality-aware  resource allocation problem for SC} as
\vspace{-1mm}\begin{equation}
\begin{aligned}
&\min_{\boldsymbol{x}\in\{0,\dots,K-1\}^P}\; J(\boldsymbol{x}) \\
&\quad\text{s.t.}\quad \mathsf{D}_q(\boldsymbol{x},\gamma)\le\delta_q,\;\;\forall q\in\mathcal{Q},
\label{eq:P}
\end{aligned}\vspace{-1mm}\end{equation}
for a given configuration $(m_b,D,\{m_k,r_k\})$ with $\epsilon_{K-1}(\gamma)\le\min_q\delta_q$. Unlike task-oriented JSCC, which transmits only features for a fixed task, \eqref{eq:P} delivers every pixel and bounds its reconstruction error by $\epsilon_{x_p}\le\delta_q/a_q(p)$. It thus guarantees the fidelity of the task-causal pixels, measured by the semantic peak SNR (S-PSNR) in Section~IV, while tolerating errors on the task-invariant pixels. \emph{Since task-causal pixels are typically few, the optimal allocation $\boldsymbol{x}$ is sparse}.
Unlike importance-aware quantization \cite{park2024vision}, which minimizes distortion under a bit budget for a single task, \eqref{eq:P} minimizes the resource cost under per-task reliability constraints, which decouple across pixels and yield the exact solution below.

\vspace{-2mm}\begin{proposition}\label{prop:thr}
Let $w_p=\max_{q\in\mathcal{Q}}a_q(p)/\delta_q$, so that $1/w_p$ is the largest error that pixel $p$ can tolerate. The unique solution of \eqref{eq:P} assigns each pixel the cheapest tier in terms of resource cost whose error is within this tolerance,
\vspace{-1mm}\begin{equation}
\begin{aligned}
x_p^\star&=\min\big\{k:\ \epsilon_k(\gamma)\le 1/w_p\big\}\\
&=\sum_{k=1}^{K-1}\boldsymbol{1}\big[w_p>\tau_k(\gamma)\big],\qquad
\tau_k(\gamma)=\frac{1}{\epsilon_{k-1}(\gamma)} .
\end{aligned}
\label{eq:sol}
\vspace{-2mm}\end{equation}
\end{proposition}
\begin{proof}
The constraints of \eqref{eq:P} hold if and only if every pixel satisfies $a_q(p)\epsilon_{x_p}\le\delta_q$ for all $q$, that is, $w_p\,\epsilon_{x_p}\le1$. The constraints therefore decouple across pixels. Since $\epsilon_k$ decreases in $k$, the feasible tiers of pixel $p$ are $\{k^\star(p),\dots,K-1\}$, which is nonempty because $w_p\le1/\min_q\delta_q$ and $\epsilon_{K-1}\le\min_q\delta_q$. Since $c_k$ increases in $k$, the minimizer is $k^\star(p)=\min\{k:w_p\epsilon_k\le1\}$, which equals the number of indices $k\ge1$ with $w_p\epsilon_{k-1}>1$, as in \eqref{eq:sol}.
\vspace{-3mm}\end{proof}

Proposition~1 shows that the \emph{optimal allocation decouples the semantic criticality of each pixel from the channel quality, and hence operates on two timescales}. The criticality-weighted causal influence $w_p$ depends only on the image and the tasks and is computed once per image, whose transmission spans many channel realizations, whereas the thresholds $\tau_k(\gamma)$ depend only on the channel, so adapting to each realization only requires re-thresholding $w_p$. The protected pixels thus correspond to the task-causal pixels $\bigcup_q\mathcal{C}_q$, and \eqref{eq:P} seeks the fewest REs for which every answer is invariant to the intervention that the base layer applies to the remaining pixels. Because each $\epsilon_k(\gamma)$ decreases with $\gamma$, fewer pixels need strong protection at higher SNR. However, solving \eqref{eq:P} requires the causal influence $a_q$, which has no closed form and evaluating it directly requires interventions on $f_q$. Finding $\mathcal{R}_q$ in Definition~1 requires a search over $2^P$ pixel subsets. Even a per-pixel estimate requires $P$ inferences per task and per image, each averaged over channel realizations. A learned estimator of $a_q$, in turn, requires labeled data for each task. We therefore \emph{seek a surrogate of $a_q$ that is obtained from a single inference and without training}.

\vspace{-2mm}\section{Proposed Solution via Zero-Shot Learning}\vspace{-1mm}
 To solve \eqref{eq:P}, Section~\ref{sec:sci} estimates the causal influence $a_q$, and hence $w_p$, without training, and Section~\ref{sec:tx} delivers each pixel on the OFDM link at the tier selected by \eqref{eq:sol}. %To solve \eqref{eq:P},  we first estimate the causal influence $a_q$ of every pixel to obtain $w_p$, and map each pixel to the tier selected by \eqref{eq:sol}. Section~\ref{sec:sci} performs the first step without training, and Section~\ref{sec:tx} realizes the tiers on the OFDM link.
\begin{algorithm}[t]
\SetAlgoLined
\DontPrintSemicolon
\caption{\small  SCI Estimation via Zero-Shot Learning \protect\footnotemark}
\label{alg:sci} \footnotesize
\KwIn{image $I$; tasks $\mathcal{Q}=\{(q,\omega_q)\}$; frozen VLM with $L$ layers}
\KwData{layer band $[\ell_0,\ell_1)=[\lfloor0.4L\rfloor,\lfloor0.7L\rfloor)$; sink $s{=}4$; blur $\sigma{=}1.2$; dilation $d{=}3$; thresholds $\boldsymbol\tau$}
\KwOut{tensor $T\in\mathbb{Z}^{H\times W\times5}$ with channels $(R,G,B,S,\mathrm{pos})$}
\BlankLine
\ForEach{$(q,\omega_q)\in\mathcal{Q}$}{
    $(\mathcal{G},\mathrm{Att})\gets\mathrm{VLM}(I,\mathrm{Prompt}(q))$\;
    $\mathcal{T}\gets$ location tokens of $\mathcal{G}$\;
    $\boldsymbol{v}\gets\max_{t\in\mathcal{T}}\ \mathrm{mean}_{\ell\in[\ell_0,\ell_1),\,\mathrm{heads}}\ \mathrm{Att}[t][\ell]\big[-1,\ \mathrm{image}\big]$\;
    $\boldsymbol{v}\gets\min\big(\boldsymbol{v},\ (s{+}1)\text{-th largest of }\boldsymbol{v}\big)$ \tcp*{sinks}
    $A_q\gets\mathrm{Norm}\big(\mathrm{Dilate}_d(\mathrm{Blur}_\sigma(\boldsymbol{v}))\big)$ \tcp*{$\approx a_q$}
}
$w\gets\mathrm{Norm}\big(\max_q\omega_qA_q\big)$ \tcp*{$\approx w_p$}
$S\gets\mathrm{UpsampleNearest}\big(1+\mathrm{Digitize}(w,\boldsymbol\tau)\big)$\;
$T[p]\gets\big(I[p],\ S[p],\ rW+c\big)$ for each pixel $p=(r,c)$\;
\Return{$T$}
\end{algorithm}
\footnotetext{\scriptsize For a task other than question answering, such as object detection or anomaly description, only $\mathrm{Prompt}(q)$ is replaced by an instruction for that task, and the tokens that state the location of the evidence form $\mathcal{T}$. The rest of the algorithm is unchanged.}
\vspace{-2mm}\subsection{Zero-Shot Learning for SCI Estimation}\label{sec:sci}\vspace{-2mm}
Algorithm~\ref{alg:sci} estimates the causal influence $a_q$ from the attention that a frozen VLM already computes. Since the VLM is frozen and the task enters only through the prompt, the estimate is zero-shot, i.e., a new task requires only a new prompt, \emph{without training, labeled data, or fine-tuning}. It returns the transmit tensor $T$, which holds, for every pixel, the three color channels, the raster position, and the SCI channel $S[p]=1+x_p^\star$, which stores the tier assigned by \eqref{eq:sol}. %We assume that the VLM answers the queries in $\mathcal{Q}$ correctly on the original image.

For each task $q$, the VLM receives the image, tokenized into an $H_m\times W_m$ grid of image tokens, and a prompt that asks for the answer to $q$ within \texttt{<answer>} tags and for the location of the evidence within \texttt{<where>} tags. Requesting the location steers the attention of the VLM toward the task-causal pixels, and the answer is generated by greedy decoding. For each generated token $t$ within the \texttt{<where>} tags, collected in $\mathcal{T}$, the algorithm takes the attention from $t$ to the image tokens and averages it over the heads and over the decoder layers $\lfloor0.4L\rfloor$ to $\lfloor0.7L\rfloor$, where localization is most reliable \cite{zhang2025mllms}. An element-wise maximum over $\mathcal{T}$ gives $\boldsymbol{v}\in\mathbb{R}^{H_m W_m}$, which assigns each image token the largest attention it receives from the location tokens.
 
A few image tokens, called attention sinks, receive large attention regardless of the task \cite{xiao2024efficient} and would otherwise dominate $\boldsymbol{v}$ after normalization. Hence, the $s$ largest values of $\boldsymbol{v}$ are capped at the $(s+1)$-th largest value. A Gaussian filter of width $\sigma$ and a grey dilation of size $d$ absorb small errors in the location of the attention peak, and a min-max normalization, denoted $\mathrm{Norm}$, sets the peak to one. This gives the SCI map $A_q\in[0,1]^{H_m\times W_m}$, whose entry for each image token estimates the causal influence $a_q(p)$ of the pixels $p$ in its patch. The maps are combined into the criticality-weighted causal influence $w=\mathrm{Norm}(\max_q\omega_qA_q)$ of Proposition~\ref{prop:thr} and quantized into tiers by $\mathrm{Digitize}$, so that $S[p]=1+x_p$. The thresholds $\boldsymbol{\tau}$ of Digitize play the role of $\tau_k$ in \eqref{eq:sol} on this normalized scale.  The SCI channel is then upsampled to the pixel grid by nearest-neighbor interpolation, which keeps region boundaries sharp.

% Replaces BOTH "TX-RX Algo" and "Tiered Transmission and Reception" (merged).
% Algorithm 3 (receiver) is removed; the last paragraph describes the receiver.
\vspace{-3mm}\subsection{Tiered Transmission and Reception}\label{sec:tx}\vspace{-2mm}
\begin{comment}
  \begin{algorithm}[t]
\SetAlgoLined\footnotesize
\DontPrintSemicolon
\caption{STAR: Transmitter}
\label{alg:tx}
\KwIn{tensor $T=(R,G,B,S,\mathrm{pos})$; base layer $(m_b,D)$; protection tiers $\{(m_k,r_k)\}_{k=1}^{K-1}$; constraint length $\kappa$}
\KwOut{symbols $\boldsymbol{s}_{\mathrm b}$ and $\{\boldsymbol{s}_k\}_{k=1}^{K-1}$}
\BlankLine
$\boldsymbol{s}_{\mathrm b}\gets\Phi_{m_b}\big(\beta(\downarrow_D(\mathrm{RGB}(T)))\big)$ \tcp*{base layer}
\For{$k=1,\dots,K-1$}{
  $\mathcal{P}_k\gets\{p:S[p]=1+k\}$\;
  $\boldsymbol{s}_k\gets\Phi_{m_k}\big(\mathcal{E}_{r_k,\kappa}(\{[\mathrm{pos}_p\,|\,R_p\,|\,G_p\,|\,B_p]\}_{p\in\mathcal{P}_k})\big)$\;
}
\Return{$\boldsymbol{s}_{\mathrm b},\ \{\boldsymbol{s}_k\}_{k=1}^{K-1}$}
\end{algorithm}
  
\end{comment}
Each tier of \eqref{eq:sol} is realized as a separate layer of the OFDM link. %, as shown in Algorithm~\ref{alg:tx}. 
The base layer decimates the image by the bilinear operator $\downarrow_D$, serializes it by the 8-bit raster map $\beta$, and maps it uncoded onto a Gray-labeled constellation $\Phi_{m_b}$, so that it consumes few REs. Each protection tier $k\ge1$ forms one self-addressed unit $[\mathrm{pos}_p\,|\,R_p\,|\,G_p\,|\,B_p]$ per pixel with $x_p=k$, encodes the units with a terminated rate-$r_k$ convolutional encoder $\mathcal{E}_{r_k,\kappa}$ of constraint length $\kappa$, and maps them onto $\Phi_{m_k}$. The parameters $D$, $m_b$, and $\{m_k,r_k\}$ are set according to the available REs. After cyclic-prefix removal and the discrete Fourier transform, the received symbol on subcarrier $n$ is
$
y[n]=h[n]\,s[n]+w[n],$
where $h[n]$ is the frequency response of a TDL-A realization with $\mathbb{E}|h[n]|^2=1$, and $w[n]$ is additive white Gaussian noise of power $N_0=10^{-(E_s/N_0)/10}$. %We use $E_s/N_0$ instead of $E_b/N_0$, because $E_s$ is fixed by the transmit power and the path loss, so all modulation orders are compared on the same physical channel.

Assuming perfect channel state information, the RX applies linear minimum mean square error (LMMSE) equalizer $\hat s[n]\!=\!h^{\ast}[n]\,y[n]/(|h[n]|^2+N_0)$. The base layer is hard-demapped, deserialized by $\beta^{-1}$, and upsampled by $D$. Each protection tier is soft-demapped into log-likelihood ratios and Viterbi-decoded, and each decoded pixel overwrites its position in the upsampled base layer. A pixel in tier $k$ is thus received with error $\epsilon_k$, as in \eqref{eq:eps}.  The $B_a$ address bits are encoded jointly with the color bits and thus share their protection, and a residual address error only misplaces a single pixel.

\begin{figure*}[t]
  \centering
  \subfloat[Fire task critical]{%
    \includegraphics[width=0.19\linewidth]{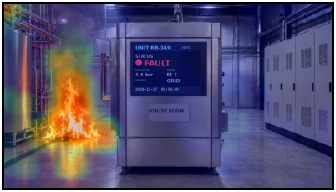}%
    \label{fig:sel_fire}}
  \hfil
  \subfloat[Room task critical]{%
    \includegraphics[width=0.19\linewidth]{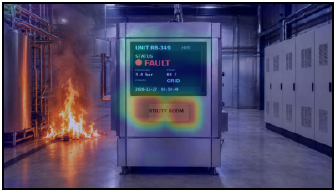}%
    \label{fig:sel_room}}
  \hfil
  \subfloat[Fire and room tasks critical]{%
    \includegraphics[width=0.19\linewidth]{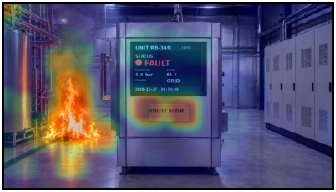}%
    \label{fig:sel_both}}
  \hfil
  \subfloat[All $7$ tasks critical]{%
    \includegraphics[width=0.19\linewidth]{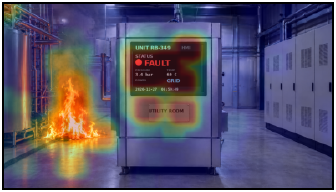}%
    \label{fig:sel_all}}
  \hfil
    \hfil
  \raisebox{-0.6ex}{\scicolorbar{0.10\linewidth}}
\vspace{-2mm}\caption{\small Continuous SCI channel overlaid on a fire-event image for different sets of critical tasks.}
  \label{fig:selective_protection}
\vspace{-6mm}\end{figure*}

\begin{figure*}[t]
  \centering
  \subfloat[Conventional, $D=4$]{%
    \includegraphics[width=0.19\linewidth]{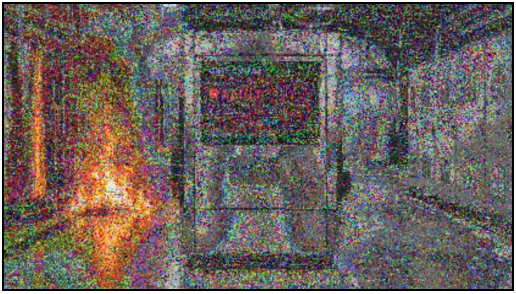}%
    \label{fig:pr_ds4}}
  \hfil
  \subfloat[Conventional, $D=1$]{%
    \includegraphics[width=0.19\linewidth]{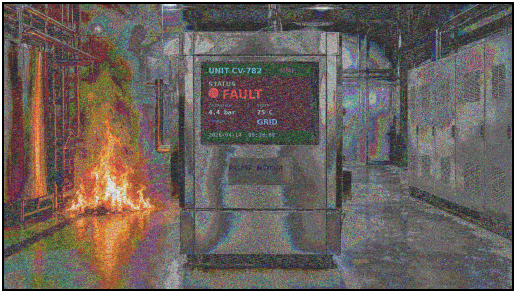}%
    \label{fig:pr_ds1}}
  \hfil
  \subfloat[STAR, room task critical]{%
    \includegraphics[width=0.19\linewidth]{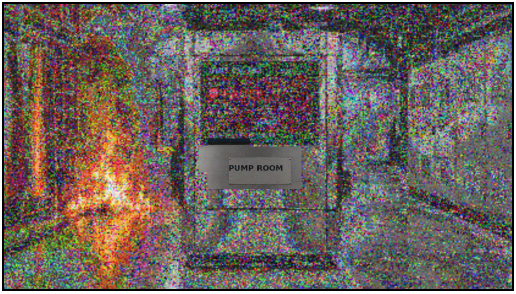}%
    \label{fig:pr_room}}
  \hfil
  \subfloat[STAR, all $7$ tasks critical]{%
    \includegraphics[width=0.19\linewidth]{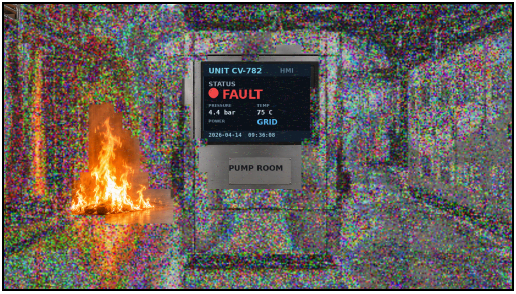}%
    \label{fig:pr_all}}
\vspace{-2mm} \caption{\small Received images of a fire-event image with $256$-QAM at $E_s/N_0=2$~dB.}
  \label{fig:recon_pumproom}
\vspace{-5mm}\end{figure*}

\vspace{-3mm}\section{Simulation Results and Analysis}\label{sec:results}\vspace{-2mm}
%\subsection{Datasets}\label{sec:datasets}\vspace{-2mm}

We evaluate STAR on a synthetic factory-monitoring dataset and a vehicle-monitoring dataset of real roadside photographs. The factory-monitoring images have $H\times W=941\times1672$ pixels and are generated from the three event templates in Fig.~\ref{fig:dataset_scenes}, with the room label and the panel readings drawn at random. We draw $20$ test images per template, which gives $|\mathcal{M}|=60$, and for ease of exposition consider the $7$ tasks of Table~\ref{tab:qa}, although STAR admits any number of tasks. Hence, each modulation order is evaluated with $420$ inferences per $E_s/N_0$ value, that is, $27{,}720$ inferences per physical layer configuration. The vehicle-monitoring dataset is the US subset of the OpenALPR benchmark~[16], with $222$ photographs and their plates as ground truth.We test on $|\mathcal{M}|=100$ images of varying resolution, with widths varying from $1280$ to $1920$ pixels, on the single task ``What is the license plate number of the vehicle?''.

\begin{table}[t]
\centering
\caption{\footnotesize Queries of the factory-monitoring dataset and sample answers.}
\vspace{-2mm}\label{tab:qa}
\renewcommand{\arraystretch}{1}\footnotesize
\begin{tabular}{p{0.58\columnwidth} p{0.30\columnwidth}}
\hline \footnotesize
\textbf{Query} & \textbf{Sample answer} \\
\hline
Which room is this? & Boiler Room \\
Equipment status? & Normal \\
Pressure reading? & $3.8$ bar \\
Temperature reading? & $50\,^\circ$C \\
Power on Grid or Backup? & Grid \\
Is there a fire? & Yes / No \\
Maintenance technician present? & Present / Absent \\
\hline
\end{tabular}
\end{table}
\vspace{-1mm}

All models run locally on one NVIDIA DGX Spark. Algorithm~\ref{alg:sci} uses a frozen Qwen2.5-VL-7B model  in bfloat16 precision. Remote inference at the control center uses Qwen3-VL-30B-A3B, a larger VLM than that of Algorithm~\ref{alg:sci}, and the judge is Qwen3-8B, both with greedy decoding.
We simulate a 5G NR OFDM link in Sionna \cite{sionna} with $30$~kHz subcarrier spacing, $14$ OFDM symbols per slot, $128$ subcarriers, and a $3.5$~GHz carrier. The grid contains no pilots and has $1792$ data REs per slot. The channel follows the 3GPP TDL-A model with a $100$~ns delay spread \cite{3gpp38901}, drawn independently for each slot. Unless stated otherwise, the STAR base layer uses uncoded 256-QAM with $D=4$, and the protection tier uses BPSK with the terminated rate-$1/3$ convolutional code of constraint length $\kappa=7$ and soft-input Viterbi decoding. Hence, $K=2$, and Digitize uses the single threshold $\tau=0.5$. A critical task has $\omega_q=2$ and any other task $\omega_q=1$, so that after normalization only the task-causal pixels of the critical tasks exceed $\tau$.

\vspace{-1mm}
We compare STAR with conventional transmission, which applies the same modulation to every pixel, either at full resolution ($D=1$) or downsampled ($D=4$), and with a robust baseline that sends the whole image with BPSK and the rate-$1/3$ code.  Downsampling serves as a stand-in for conventional compression, because standard JPEG fails abruptly over an uncoded link at low $E_s/N_0$, where bit errors propagate through its variable-length code, whereas downsampling degrades gracefully. We also compare with a hybrid DeepJSCC, in which the convolutional DeepJSCC of \cite{bourtsoulatze2019deep} maps the image to $0.79$\,M complex symbols per factory image, placed directly on the same OFDM grid, and a CNN on the received symbols answers the 7 tasks of Table~I with classification heads. It  delivers both a reconstructed image and the answers without any VLM. The code is available at \url{https://github.com/THE-TRAIN-LAB/STAR-SemCom}.

\vspace{-2mm}\subsection{Evaluation Metrics}\label{sec:metrics}\vspace{-2mm}
To model remote inference at the control center, a VLM answers each query on the received image, and a judge large language model (LLM) decides whether the answer is semantically equivalent to the ground truth \cite{zheng2023judging}. %, since a correct answer may be paraphrased differently from the ground truth, such as ``50 degrees Celsius'' for ``$50\,^\circ$C''. 
Let $z_{i,q}=1$ if the judge accepts the answer for image $i\in\mathcal{M}$ and task $q\in\mathcal{Q}$, and $z_{i,q}=0$ otherwise. The semantic accuracy (S-Acc) is
$
\text{S-Acc}=\frac{1}{|\mathcal{M}|\,|\mathcal{Q}|}\sum_{i\in\mathcal{M}}\sum_{q\in\mathcal{Q}} z_{i,q},$
and $\text{S-Acc}_q$ denotes the S-Acc of task $q$ alone.

To verify the stored data, the evidence for each answer must also be delivered at high quality. Let $\mathcal{C}_{\mathrm{gt}}$ denote the ground-truth task-causal region of an image $I$, given by the generator for the factory-monitoring dataset and by the plate annotation for the vehicle-monitoring dataset, and let $\hat I$ be the received image. The S-PSNR quantifies the quality of evidence and is defined as
\vspace{-3mm}\begin{equation} \small
\text{S-PSNR}=10\log_{10}\frac{3|\mathcal{C}_{\mathrm{gt}}|\,(2^b-1)^2}{\sum_{p\in\mathcal{C}_{\mathrm{gt}}}\sum_{c\in\{R,G,B\}}\big(\hat I[p,c]-I[p,c]\big)^2},
\label{eq:spsnr}
\vspace{-1mm}\end{equation}
where $c$ indexes the color channels. We report $99$~dB when $\hat I=I$ on $\mathcal{C}_{\mathrm{gt}}$ and average the S-PSNR over $\mathcal{M}$.
 A received image that attains both a high S-Acc and a high S-PSNR attains high \emph{semantic reliability}.

% ==========================================================================
\vspace{-2mm}\subsection{Analysis using Factory-Monitoring Dataset}\label{sec:res_factory}\vspace{-1mm}

Fig.~\ref{fig:selective_protection} shows the SCI channel overlaid on a sample fire-event image for different sets of critical tasks. The protected region covers exactly the flame for the fire task (Fig.~2a), the room nameplate for the room task (Fig.~2b), and both regions for both tasks (Fig.~2c). When all $7$ tasks are critical (Fig.~2d), it additionally covers the machine status panel. In every case, the task-causal pixels form a small fraction of the image, whereas most pixels are task-invariant. The operator thus steers the protection by changing the critical tasks in the prompt, without any training.

%Fig.~\ref{fig:selective_protection} shows the SCI channel of the fire-event image for different critical tasks, wherein the operator  steers the protection by changing the critical tasks, without any training.
Fig.~\ref{fig:recon_pumproom} shows received images with $256$-QAM at $E_s/N_0=2$~dB. The conventional scheme corrupts the text both at $D=4$ (Fig.~\ref{fig:pr_ds4}) and at $D=1$ (Fig.~\ref{fig:pr_ds1}), although the latter uses $16$ times more REs, so it can only trade semantic reliability against REs. On the other hand, STAR delivers the room nameplate when only the room task is critical (Fig.~\ref{fig:pr_room}) and every task-causal region when all $7$ tasks are critical (Fig.~\ref{fig:pr_all}).

\begin{figure}[t]
  \centering
  \vspace{-1mm}
  \includegraphics[width=\columnwidth]{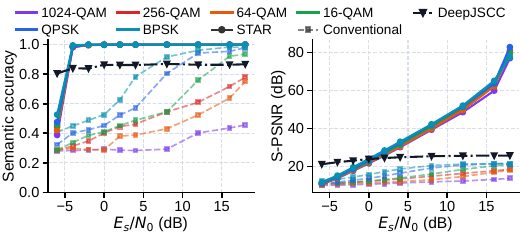}
  \vspace{-7mm}\caption{\small Performance at $D=4$ on the factory-monitoring dataset.}
  \label{fig:ds4}
\vspace{-3mm}\end{figure}

\begin{figure}[t]
  \centering
  \vspace{-3mm}
  \includegraphics[width=\columnwidth]{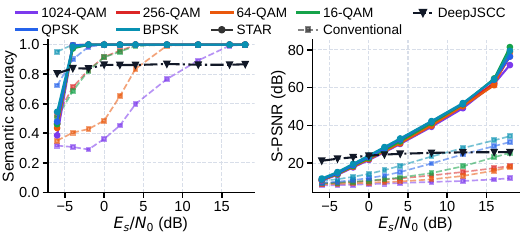}
  \vspace{-6mm}\caption{\small STAR at $D=4$, the conventional scheme at $D=1$, and the DeepJSCC on the factory-monitoring dataset.}
  \label{fig:ds1}
\vspace{-1mm}\end{figure}

% \begin{figure}[t]
%   \centering
%  \vspace{-1mm} {%
%     \includegraphics[width=0.48\columnwidth]{STAR_vs_QAM_semantic_accuracy_DS4_compressed.pdf}%
%     \label{fig:ds4_semacc}}
%   \hfil
%  \vspace{-1mm} {%
%     \includegraphics[width=0.48\columnwidth]{STAR_vs_QAM_S-PSNR_DS4_compressed.pdf}%
%     \label{fig:ds4_spsnr}}
%   \vspace{-2mm}\caption{\small STAR and the conventional scheme at $D=4$ on the factory-monitoring dataset.}
%   \label{fig:ds4}
% \vspace{-3mm}\end{figure}

% \begin{figure}[t]
%   \centering
%  \vspace{-1mm} {%
%     \includegraphics[width=0.48\columnwidth]{STAR_vs_QAM_semantic_accuracy_DS1_no_compression.pdf}%
%     \label{fig:ds1_semacc}}
%   \hfil
%  \vspace{-1mm} {%
%     \includegraphics[width=0.48\columnwidth]{STAR_vs_QAM_S-PSNR_DS1_no_compression.pdf}%
%     \label{fig:ds1_spsnr}}
%  \vspace{-2mm} \caption{\small STAR and the conventional scheme at $D=1$ on the factory-monitoring dataset.}
%   \label{fig:ds1}
% \vspace{-3mm}\end{figure}

% Fig.~\ref{fig:ds4} and Fig.~\ref{fig:ds1} compare STAR with the conventional scheme at $D=4$ and $D=1$, with all $7$ tasks critical and the same modulation order, from BPSK to $1024$-QAM, for both schemes. STAR attains $\text{S-Acc}\ge0.99$ for $E_s/N_0\ge-4$~dB with every modulation order. The conventional scheme never exceeds $0.98$ at $D=4$, and at $D=1$ it needs $4$~dB with $256$-QAM and $16$~dB with $1024$-QAM to reach $0.99$. Uncoded BPSK and QPSK at $D=1$ attain a comparable S-Acc, but their S-PSNR stays below $35$~dB, whereas that of STAR reaches $76.9$~dB at $18$~dB with a $256$-QAM base layer.

Fig.~\ref{fig:ds4} and Fig.~\ref{fig:ds1} compare STAR with the conventional scheme at $D=4$ and $D=1$, with all $7$ tasks critical and the same modulation order, from BPSK to $1024$-QAM, for both schemes. STAR attains S-Acc $\geq 0.98$ for $E_s/N_0 \geq -4$~dB with every modulation order. The conventional scheme never reaches $0.99$ at $D=4$, and with $256$-QAM it stays below $0.78$, whereas at $D=1$ it needs $4$~dB with $256$-QAM and $16$~dB with $1024$-QAM to reach $0.99$. Uncoded BPSK and QPSK at $D=1$ attain a comparable S-Acc, but their S-PSNR stays below $35$~dB, whereas that of STAR reaches $76.9$~dB at $18$~dB with a $256$-QAM base layer. The hybrid DeepJSCC degrades gracefully at low $E_s/N_0$ but saturates at an S-Acc of $0.87$. Its lossy reconstruction also saturates at an S-PSNR of about $26$\,dB, so the exact values of the task-causal pixels are not delivered. Moreover, its classification heads answer only the tasks they were trained on, so a new task or the vehicle-monitoring domain requires new labeled data and retraining, whereas STAR only changes the prompt.

%Fig.~\ref{fig:ds4} and Fig.~\ref{fig:ds1} compare STAR with the conventional scheme at $D=4$ and $D=1$, with all $7$ tasks critical and the same modulation order, from BPSK to $1024$-QAM, for both schemes. STAR attains $\text{S-Acc}\ge0.98$ from $E_s/N_0=-4$~dB for every modulation order. The conventional scheme never exceeds $\text{S-Acc}=0.98$ at $D=4$, and at $D=1$ it needs $4$~dB with $256$-QAM and $16$~dB with $1024$-QAM to reach $0.99$. Uncoded BPSK and QPSK at $D=1$ reach an S-Acc comparable to STAR, but their S-PSNR stays below $35$~dB, whereas that of STAR reaches $76.9$~dB at $18$~dB with a $256$-QAM base layer. STAR therefore attains the highest semantic reliability at every $E_s/N_0$.

\begin{figure}[t]
  \vspace{-3mm}\centering
  \vspace{-2mm}\subfloat[]{%
    \includegraphics[width=0.48\columnwidth]{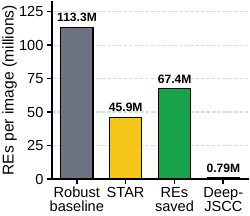}%
    \label{fig:resource_saving}}
  \hfil
   \vspace{-2mm}\subfloat[]{%
    \includegraphics[width=0.48\columnwidth]{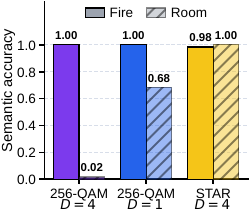}%
    \label{fig:selective_2db}}
\vspace{-2mm}  \caption{\small Factory-monitoring dataset. (a) REs per image of STAR, and the baselines. (b) Per-task S-Acc with $256$-QAM at $E_s/N_0=2$~dB, with only the room task critical for STAR.}
  \label{fig:factory_bars}
\vspace{-3mm}\end{figure}
% \begin{figure}[t]
%  \vspace{-5mm} \centering
%   \subfloat[\scriptsize Conventional, $D\!=\!1$]{%
%     \includegraphics[width=0.29\linewidth]{plate_panel_10dB_conv_256QAM_DS1.pdf}\label{fig:pl_d1}}
%   \hfil
%   \subfloat[\scriptsize Conventional, $D\!=\!4$]{%
%     \includegraphics[width=0.29\linewidth]{plate_panel_10dB_conv_256QAM_DS4.pdf}\label{fig:pl_d4}}
%   \hfil
%   \subfloat[\footnotesize STAR]{%
%     \includegraphics[width=0.29\linewidth]{plate_panel_10dB_STAR_256QAM_DS4.pdf}\label{fig:pl_star}}
  
%   \vspace{-2mm}\caption{\small Vehicle-monitoring dataset. (a)--(c) Received images of a vehicle image with $256$-QAM at $E_s/N_0=10$~dB. }
%   \label{fig:plate_panels}
% \vspace{-1mm}\end{figure}

\begin{figure}[t]
 \vspace{-2mm}\centering
  \includegraphics[width=\columnwidth]{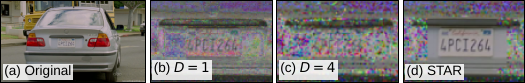}
  \vspace{-3mm}\caption{\small Vehicle-monitoring dataset, $256$-QAM at $E_s/N_0\!\!=\!\!10$~dB. (a) Original image. (b)--(d) Received image zoomed in to the number plate.}
  \label{fig:plate_panels}
\vspace{-1mm}\end{figure}

% \begin{figure}[t]
%  \vspace{-4mm} \centering
%   \subfloat[]{%
%     \vspace{-1mm}\includegraphics[width=0.48\columnwidth]{plate_accuracy_vs_EsN0.pdf}%
%     \label{fig:plate_acc}}
%   \hfil
%   \subfloat[]{%
%     \vspace{-1mm}\includegraphics[width=0.48\columnwidth]{plate_S-PSNR_vs_EsN0.pdf}%
%     \label{fig:plate_spsnr}}
%  \vspace{-3mm} \caption{\small S-Acc and S-PSNR of the plate task. The dashed line in (a) is the S-Acc of the speaker on the original images.}
%   \label{fig:plate_curves}
% \end{figure}

\begin{figure}[t]
  \vspace{-2mm}\centering
  \includegraphics[width=\columnwidth]{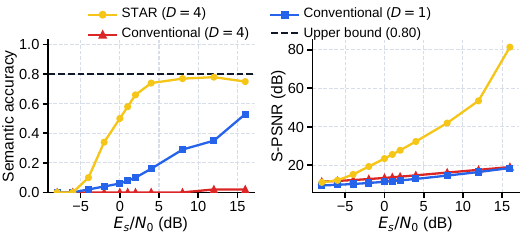}
  \vspace{-6mm}\caption{\small S-Acc and S-PSNR of the plate task with 256-QAM. The upper bound is the S-Acc attained on the images at the transmitter.}
  \label{fig:plate_curves}
\vspace{-4mm}\end{figure}
\begin{figure}[t]
  \centering
\includegraphics[height=0.45\linewidth]{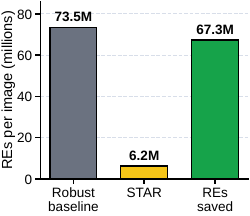}
   \vspace{-3mm} \caption{\small  REs per image of the robust baseline and STAR.}
  \label{fig:plate_res}
\vspace{-2mm}\end{figure}

For a comparison of REs at comparable semantic reliability, the robust baseline applies the protection of STAR to every pixel. As shown in Fig.~\ref{fig:resource_saving}, it uses $113.3$~million REs per image, whereas STAR uses $45.9$~million, a reduction of $59.5\%$. The hybrid DeepJSCC uses only $0.79$ million REs, at the cost of the lower S-Acc and S-PSNR in Figs.~4 and~5 and of task-specific training.
%A fair comparison of REs requires comparable semantic reliability. We therefore compare STAR with the robust reference, which transmits the whole raw image with BPSK and the rate-$1/3$ convolutional code. Its S-PSNR and S-Acc are at least those of STAR, since its critical pixels use the same modulation and coding and its normal pixels are not downsampled. As shown in Fig.~\ref{fig:resource_saving}, the robust reference uses $113.3$~million REs per image, whereas STAR with a $256$-QAM base layer uses $45.9$~million, which is a reduction of $59.5\%$.
The protection can also target only the information that the conventional schemes lose. Fig.~\ref{fig:selective_2db} shows the per-task S-Acc at the operating point of Fig.~\ref{fig:recon_pumproom}. The fire occupies a large region and attains $\text{S-Acc}_q=1.00$ with both conventional schemes, whereas the small room nameplate attains only $0.02$ at $D=4$ and $0.68$ at $D=1$. However, responding to a fire requires knowing its room. With only the room task critical, STAR attains $1.00$ on the room task and $0.98$ on the fire task.

%The critical tasks can also be chosen to protect only the information that the conventional schemes lose. Fig.~\ref{fig:selective_2db} shows the per-task S-Acc at the operating point of Fig.~\ref{fig:recon_pumproom}. The fire occupies a large region, so the fire task attains $\text{S-Acc}_q=1.00$ with both conventional schemes. The room nameplate occupies a small region, and the room task attains only $0.02$ at $D=4$ and $0.68$ at $D=1$, although the control center cannot respond to a fire without knowing its room. With only the room task critical, STAR attains $1.00$ on the room task and $0.98$ on the fire task.

\vspace{-4mm}\subsection{Domain-Agnostic Operation: Vehicle-Monitoring Dataset}\label{sec:domain}\vspace{-2mm}
In vehicle monitoring, such as traffic enforcement, any action requires identifying the vehicle, which depends on its license plate, a small region of the image.
STAR is applied to vehicle monitoring with no change to any model, parameter, or the physical layer, only the prompt. At $E_s/N_0=10$\,dB with 256-QAM, the plate is corrupted by the conventional scheme at $D=1$ and $D=4$ (Figs.~7b, 7c) but remains readable with STAR as Fig.~7d shows.
Fig.~\ref{fig:plate_curves} shows that STAR reaches $\text{S-Acc}=0.74$ at $4$~dB and $0.78$ at $12$~dB, whereas the conventional scheme reaches only $0.53$ at $16$~dB with $D=1$ and stays at or below $0.02$ with $D=4$. At $16$~dB, the S-PSNR of STAR is $81.7$~dB, whereas both conventional schemes remain below $19.5$~dB. Fig.~\ref{fig:plate_res} shows that at comparable semantic reliability, the robust baseline uses $73.5$~million REs per image and STAR $6.2$~million, a reduction of $91.5\%$, which exceeds that of the factory-monitoring dataset because the task-causal pixels are sparser.

%Fig.~\ref{fig:plate_curves} reports the S-Acc and S-PSNR of the plate task with $256$-QAM.The dashed line at $\text{S-Acc}=0.80$ is the S-Acc of the speaker over an error-free channel, which is the upper bound for every scheme. STAR reaches $\text{S-Acc}=0.74$ at $4$~dB and $0.78$ at $12$~dB, whereas the conventional scheme reaches only $0.53$ at $16$~dB at $D=1$ and stays at or below $0.02$ at $D=4$. At $16$~dB, the S-PSNR of STAR is $81.7$~dB, whereas both conventional schemes remain below $19.5$~dB. At comparable semantic reliability, the robust reference uses $73.5$~million REs per image, whereas STAR uses $6.2$~million, as shown in Fig.~\ref{fig:plate_re}. This is a reduction of $91.5\%$, which is larger than on the factory-monitoring dataset because the critical pixels are sparser.

%\vspace{-3mm}\subsection{Computation Time}\label{sec:res_time}\vspace{-2mm}
The computation time of STAR is dominated by SCI estimation, which takes at most $5.3$~s per task on one GPU with sequential inference. This latency can be reduced by parallel task processing and quantized VLMs, and we relegate the evaluation of the end-to-end latency to future work. Moreover, we will extend STAR to more than two protection tiers and evaluate it on larger and more complex datasets.
%The computation time of STAR is dominated by the SCI channel generator. With our experimental stack, which uses eager attention in bfloat16 without batching, the generator takes $5.3$~s per task. For real-world deployment, this time can be reduced substantially with inference tools such as TensorRT-LLM and model quantization.

% ==========================================================================
\vspace{-2mm}\section{Conclusion}\label{sec:conclusion}\vspace{-2mm}
We proposed STAR, a training-free framework that protects the image regions on which the tasks of a RX depend while transmitting the full image. STAR derived the SCI from the decoder attention of a frozen VLM and protects the task-causal pixels with a coded layer over an unchanged OFDM link. At semantic reliability comparable to the robust reference, STAR reduced the REs per image by $59.5\%$ on the factory-monitoring dataset and by $91.5\%$ on the vehicle-monitoring dataset, without any change to the models. %Future work will extend the SCI to more than two protection tiers and will adapt the threshold to $E_s/N_0$.

\vspace{-2mm}\bibliographystyle{IEEEbib}
\def\baselinestretch{0.82}
\bibliography{refs}

\end{document}